\documentclass[useAMS,usenatbib,usegraphicx]{mn2e}
\usepackage{epsf}

 \newcommand{\bm}[1]{\mbox{\boldmath$#1$}}

\newcommand{\simgt}{\lower.5ex\hbox{$\; \buildrel > \over \sim \;$}}
\newcommand{\simlt}{\lower.5ex\hbox{$\; \buildrel < \over \sim \;$}}

\title[The complex cluster SDSS J1029+2623]
{The Hidden Fortress: Structure and substructure of 
the complex strong lensing cluster SDSS J1029+2623
}

\author[M.~Oguri et al.]
{Masamune Oguri,$^1$\thanks{E-mail: masamune.oguri@ipmu.jp} 
Tim Schrabback,$^{2,3}$
Eric Jullo,$^4$
Naomi Ota,$^5$\newauthor
Christopher S. Kochanek,$^6$
Xinyu Dai,$^7$
Eran O. Ofek,$^8$
Gordon T. Richards,$^9$\newauthor
Roger D. Blandford,$^3$ 
Emilio E. Falco$^{10}$ and
Janine Fohlmeister$^{11}$\\
$^1$Kavli Institute for the Physics and Mathematics of the Universe
(Kavli IPMU, WPI), University of Tokyo, Chiba 277-8583, Japan\\ 
$^2$Argelander-Institut f\"{u}r Astronomie, Auf dem H\"{u}gel 71,
53121 Bonn, Germany\\
$^3$Kavli Institute for Particle Astrophysics and Cosmology, Stanford
University, 382 Via Pueblo Mall, Stanford, CA 94305-4060, USA\\ 
$^4$Aix Marseille Universit\'{e}, CNRS, LAM (Laboratoire d'Astrophysique
de Marseille) UMR 7326, 13388, Marseille, France\\
$^5$Department of Physics, Nara Women's University,
Kitauoyanishimachi, Nara, Nara 630-8506, Japan\\
$^6$Department of Astronomy, The Ohio State University, Columbus, OH
43210, USA\\
$^7$Homer L. Dodge Department of Physics and Astronomy, University of
Oklahoma, Norman, OK 73019, USA\\
$^{8}$Benoziyo Center for Astrophysics, Weizmann Institute of
Science, 76100 Rehovot, Israel\\
$^{9}$Department of Physics, Drexel University, 3141 Chestnut Street,
Philadelphia, PA 19104, USA\\
$^{10}$Harvard-Smithsonian Center for Astrophysics, Cambridge, MA02138, USA\\
$^{11}$Astronomisches Rechen-Institut, Zentrum f\"{u}r Astronomie der
Universit\"{a}t Heidelberg, M\"{o}nchofstr. 12-14, 69120 Heidelberg, Germany
} 

\begin{document}

\date{\today}

\voffset- .5in

\pagerange{\pageref{firstpage}--\pageref{lastpage}} \pubyear{}

\maketitle

\label{firstpage}

\begin{abstract}
We present {\it Hubble Space Telescope (HST)} Advanced Camera for
Surveys (ACS) and Wide Field Camera 3 (WFC3) observations of 
SDSS~J1029+2623, a three-image quasar lens system 
produced by a foreground cluster at $z=0.584$. Our strong lensing
analysis reveals 6 additional multiply imaged galaxies in addition to
the multiply imaged quasar. We confirm the complex nature of the mass
distribution of the lensing cluster, with a bimodal dark matter
distribution which deviates from the {\it Chandra} X-ray surface
brightness distribution. The Einstein radius of the lensing cluster is
estimated to be $\theta_{\rm E}=15\farcs2\pm0\farcs5$ for the 
quasar redshift of $z=2.197$. We derive a radial mass distribution
from the combination of strong lensing, {\it HST}/ACS weak lensing,
and Subaru/Suprime-cam weak lensing analysis results, 
finding a best-fit virial mass of $M_{\rm vir}=1.55^{+0.40}_{-0.35} 
\times 10^{14}h^{-1}M_\odot$ and a concentration parameter of 
$c_{\rm vir}=25.7^{+14.1}_{-7.5}$. The lensing mass estimate at the outer 
radius is smaller than the X-ray mass estimate by a factor of 
$\sim 2$. We ascribe this large mass discrepancy to shock 
heating of the intracluster gas during a merger, which is also
suggested by the  complex mass and gas distributions and the high
value of the concentration parameter. In the {\it HST} image, we
also identify a probable galaxy, GX, in the vicinity of the faintest
quasar image C. In strong lens models, the inclusion of GX explains the
anomalous flux ratios between the quasar images. The morphology of the
highly elongated quasar host galaxy is also well reproduced. The
best-fit model suggests large total magnifications of $30$ for the
quasar and $35$ for the quasar host galaxy, and has an AB time delay 
consistent with the measured value.
\end{abstract}


\section{Introduction}
\label{sec:intro}

SDSS~J1029+2623 \citep{inada06,oguri08} is a unique quasar lens system
with three images of the $z_s=2.197$ quasar produced by a foreground
cluster of galaxies at $z_l=0.584$. Such ``naked cusp'' configurations
are predicted to comprise a significant fraction of cluster-scale
quasar lenses \citep{oguri04,li07,minor08}. All three quasar images
are also detected at radio \citep{kratzer11} and X-ray wavelengths
\citep{ota12}, and show mini-broad absorption line (mini-BAL) 
features in the Ly$\alpha$, SiIV, and CIV emission lines. 

These earlier observations already implied that the lensing cluster is
likely complex. The optical images clearly show two bright central
galaxies in the lensing cluster. Follow-up spectroscopy found that the
two galaxies have a velocity difference of 
$\Delta v\simeq 2800~{\rm km\,s^{-1}}$ \citep{oguri08}, which might
imply a high-speed collision of two massive haloes. {\it Chandra}
X-ray observations show that the X-ray surface mass distribution is
centred on one of the central galaxies, with an additional X-ray
subpeak $\sim 9''$  North-West of the main peak that has no optical 
counterpart \citep{ota12}. This complex X-ray morphology also
suggests that the lensing cluster is in a merger. It is crucial to
construct a reliable mass model from a lensing analysis in order to
draw any conclusion on the dynamical state of the cluster. 

Another interesting feature of this system is the anomalous flux
ratios seen in the quasar images. Simple mass models predict the
relative brightnesses of the three quasar images $A$, $B$, and $C$
should be $B\approx C > A$, whereas optical, radio, and X-ray
observations all found relative brightnesses of $B\approx A >  C$. 
The fact that this ``anomaly'' is seen in radio \citep{kratzer11} and
absorption-corrected X-ray \citep{ota12} flux 
ratios indicates that it is not due to extinction by any intervening
matter. In addition, the rough agreement of the flux ratios between
three different wavelengths implies that it is not entirely due to
stellar microlensing  \citep[but see also][]{motta12}, particularly
since microlensing generally leads to large differences between
optical and X-ray flux ratios \cite[e.g.,][]{dai10a}. Thus the most
plausible explanation for the anomalous flux ratio is millilensing by
(dark halo) substructures \citep{kratzer11}. An accurate macro lens
model is crucial for breaking the degeneracy between the macro model
uncertainties and the amount of substructure required to account for
the flux ratios.   

In this paper, we present a new lensing analysis of SDSS~J1029+2623
utilizing deep multiband {\it Hubble Space Telescope (HST)}
observations of this system. Several multiply imaged galaxies, which
are newly identified from the {\it HST} images, are used to construct
an accurate mass model of the core of the lensing cluster. Strong
lensing results are then combined with weak lensing results from the
{\it HST} images and our earlier wide-field weak lensing analysis
based on the Subaru/Suprime-cam images \citep{oguri12}. We then use
the mass model to interpret the {\it Chandra} X-ray surface brightness
profiles \citep{ota12} and the anomalous flux ratios between the
quasar images. 

The structure of this paper is as follows. After briefly describing
the observations in Section~\ref{sec:obs}, we present the strong
lensing analysis in Section~\ref{sec:slens} and the weak lensing
analysis in Section~\ref{sec:wlens}. We then combine the various
observations in Section~\ref{sec:comb} to discuss the properties of
the lensing cluster. Section~\ref{sec:flux} is devoted to discussions
of the anomalous flux ratio between the quasar images. Finally we
summarize the results on Section~\ref{sec:conclusion}.
Throughout this paper we assume a flat cosmological model with matter
density $\Omega_M=0.275$, cosmological constant
$\Omega_\Lambda=0.725$, and dimensionless Hubble constant $h=0.702$
\citep{komatsu11}. All photometry is in the AB system without
correcting for Galactic extinction.  
 
\section{Observations}
\label{sec:obs}

\begin{figure}
\begin{center}
 \includegraphics[width=0.99\hsize]{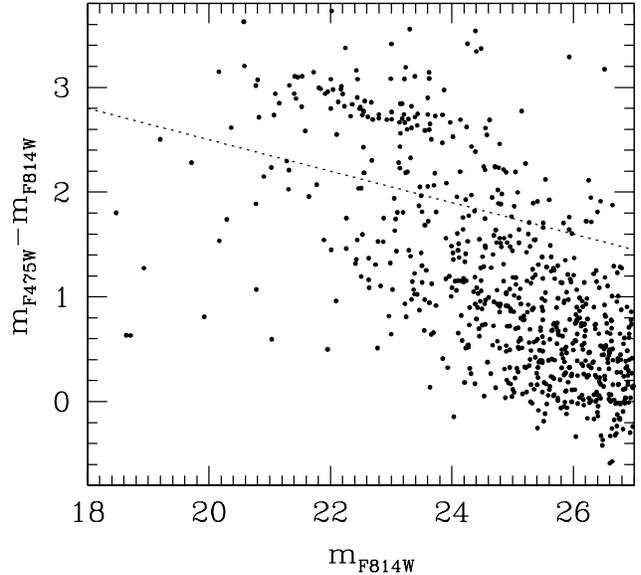}
\end{center}
\caption{The colour-magnitude diagram from the {\it HST} ACS images. 
Here the magnitudes are $1''$ diameter aperture magnitudes. The dotted
line indicates the colour cut used to eliminate cluster member
galaxies for weak lensing analysis presented in
Section~\ref{sec:wlens}. 
\label{fig:colmag}}
\end{figure}

We observed SDSS~J1029+2623 with the Advanced Camera for Surveys (ACS)
and the Wide Field Camera 3 (WFC3) on {\it HST} (GO-12195; PI Oguri).
The observation consisted of 2 orbits (total exposure 5276~sec) of
ACS/F475W imaging on 2011 May 15, 3 orbits (total exposure 8010~sec)
of ACS/F814W imaging on 2011 April 17, and 2 orbits (total exposure
5223.5~sec) of WFC3/F160W imaging on 2011 April 15. We mostly follow
the data reduction procedure described in \citet{schrabback10}, which
uses MultiDrizzle \citep{koekemoer02} for cosmic ray removal,
distortion removal and stacking. A pixel scale of $0\farcs05$ is
adopted for both the ACS and WFC3 images. The ACS (WFC3) frames are
drizzled using the lanczos3 (square) kernel, with a {\tt pixfrac} of
1 (0.8). The ACS data are preprocessed using the pipeline CALACS,
which also removes bias striping and provides a pixel-level charge
transfer inefficiency correction.   

We use SExtractor \citep{bertin96} to construct a photometric
catalogue of the galaxies in the {\it HST} images. For each 
galaxy, we derive a total magnitude from the {\tt MAG\_AUTO} parameter 
and a $1''$ diameter aperture magnitude. We show the colour-magnitude
diagram of the SDSS~J1029+2623 field from the two ACS images in
Figure~\ref{fig:colmag}. The Figure shows a clear red-sequence of
passively evolving galaxies in the lensing cluster at 
$m_{\rm F475W}-m_{\rm F814W}\sim 3$.   
  
We also use wide-field Subaru Suprime-cam \citep{miyazaki02} 
images of SDSS~J1029+2623 for the analysis. These images consist of
1200~sec $g$-band, 2700~sec $r$-band, and 1920~sec $i$-band images,
with a seeing FWHM of $0\farcs65$ in $r$-band. The weak lensing
analysis of the Subaru data was presented in \citet{oguri12}, which we
adopt in this paper.  

SDSS~J1029+2623 was also observed with {\it Chandra} X-ray
observatory (ObsID 11755; PI Oguri) on 2010 March 11 for 60~ksec. The
X-ray image clearly reveals X-ray emission both from the lensing
cluster and the three quasar images. The data 
reduction and basic spectral and image analysis were detailed in
\citet{ota12}. 

\section{Strong lensing analysis}
\label{sec:slens}

\begin{figure*}
\begin{center}
 \includegraphics[width=0.99\hsize]{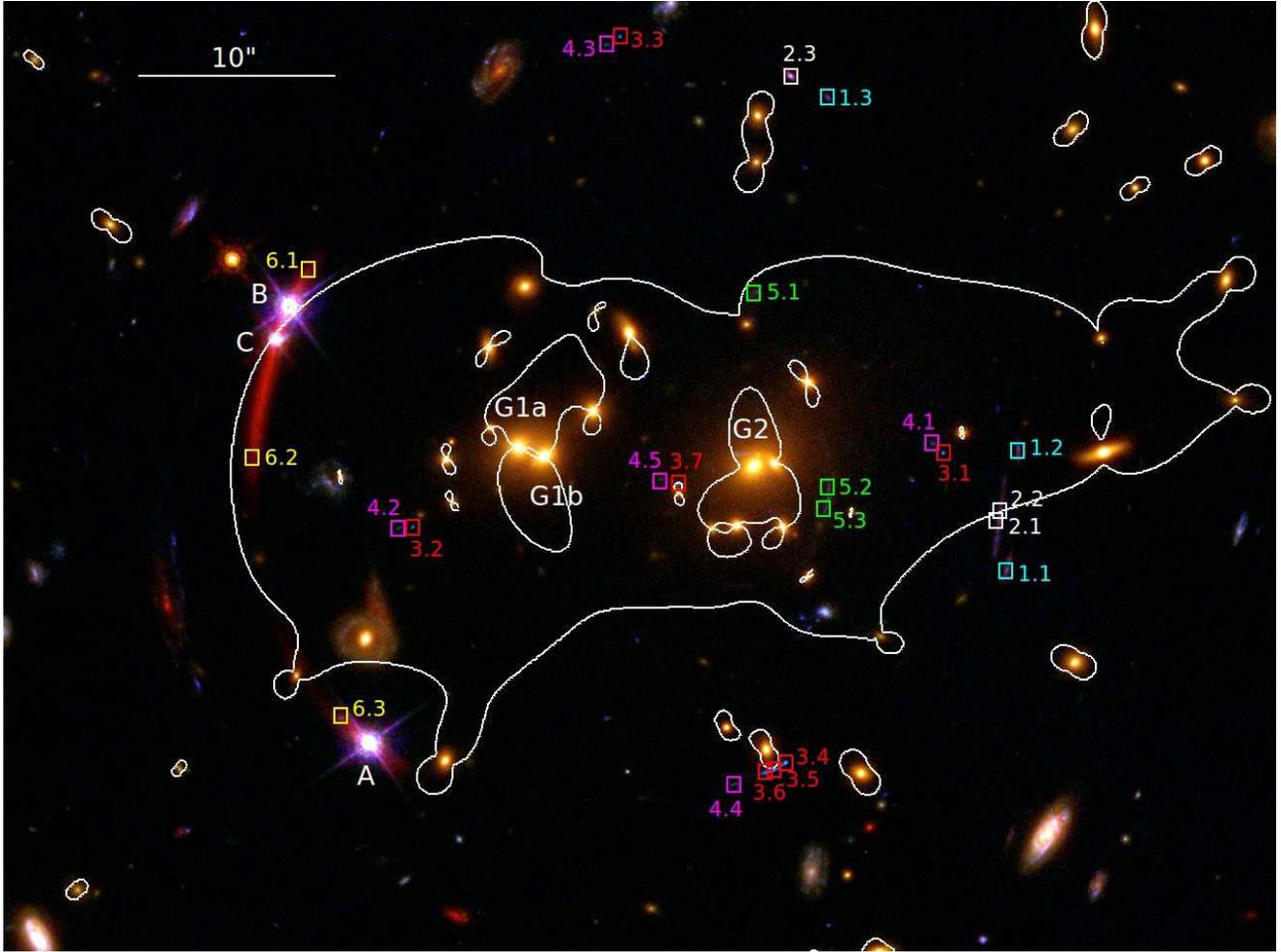}
\end{center}
\caption{Colour image of the SDSS~J1029+2623 field from the {\it HST}
  ACS/F475W, ACS/F814W, and WFC3/F160W observations. North is up and
  West is to the right. The three quasar images are labelled by A, B,
  and C. Squares with ID numbers indicate the locations of identified
  multiple images (see also Table~\ref{tab:img}). Galaxies G1a, G1b, G2
  are the central galaxies of the lensing cluster. Solid lines show
  the critical lines predicted by our best-fit mass model at
  $z_s=2.197$, the redshift of the strongly lensed quasar. 
\label{fig:sdss1029rgb_all}}
\end{figure*}

\subsection{Modelling method}

\begin{figure*}
\begin{center}
 \includegraphics[width=0.95\hsize]{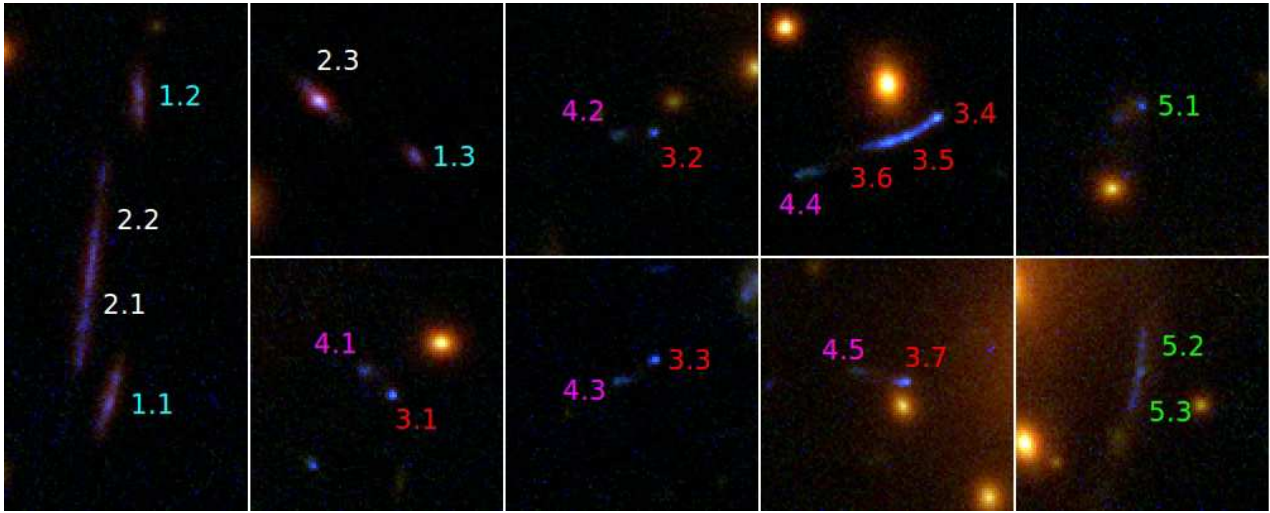}
\end{center}
\caption{Similar to Figure~\ref{fig:sdss1029rgb_all}, but zooming in
  on the multiply imaged galaxies identified in this paper.
\label{fig:zoom_all}}
\end{figure*}

The main objective of the {\it HST} imaging observation was to
constrain the central mass distribution of the lensing cluster using
strong lensing information. For this purpose, we identify multiply
imaged galaxies in addition to the three quasar images based on their
colours and morphologies combined with matching multiple image
candidates while iteratively refining the mass models (see below for
details of our mass model). In total, we identified 5 additional sets
of multiply imaged galaxies (ID 1-5), which are shown in
Figures~\ref{fig:sdss1029rgb_all} and \ref{fig:zoom_all}, and
summarized in Table~\ref{tab:img}. In addition, the {\it HST} image
shown in Figure~\ref{fig:sdss1029rgb_all} exhibits a prominent host
galaxy of the lensed quasar which is highly elongated due to 
lensing. In order to take account of the shape of the lensed host
galaxy, we include an additional set of multiple images (ID 6) which
roughly correspond to the edge of the host galaxy. To summarize, in
this paper we use the positions of 27 multiple images of 7 systems as
constraints.  

\begin{table}
 \caption{List of multiple images for strong lensing analysis.
\label{tab:img}}    
 \begin{tabular}{@{}cccc}
 \hline
   ID
   &  R.A. (J2000)$^a$
   &  Decl. (J2000)$^a$
   &  redshift$^b$ \\
 \hline
A   & 157.308139 & 26.388441 & 2.197 \\
B   & 157.309398 & 26.394600 & \\
C   & 157.309612 & 26.394125 & \\
1.1 & 157.298104 & 26.390875 & [1.93--2.36]\\
1.2 & 157.297918 & 26.392583 & \\
1.3 & 157.300902 & 26.397536 & \\
2.1 & 157.298228 & 26.391494 & [1.84--2.27]\\
2.2 & 157.298197 & 26.391783 & \\
2.3 & 157.301516 & 26.397847 & \\
3.1 & 157.299103 & 26.392536 & [2.62--3.24]\\
3.2 & 157.307476 & 26.391480 & \\
3.3 & 157.304173 & 26.398397 & \\
3.4 & 157.301584 & 26.388161 & \\
3.5 & 157.301770 & 26.388063 & \\
3.6 & 157.301878 & 26.388022 & \\
3.7 & 157.303243 & 26.392097 & \\
4.1 & 157.299258 & 26.392661 & [2.58--3.26]\\
4.2 & 157.307677 & 26.391452 & \\
4.3 & 157.304390 & 26.398286 & \\
4.4 & 157.302359 & 26.387869 & \\
4.5 & 157.303531 & 26.392152 & \\
5.1 & 157.302080 & 26.394772 & [0.89--1.02]\\
5.2 & 157.300917 & 26.392008 & \\
5.3 & 157.300964 & 26.391730 & \\
6.1 & 157.309073 & 26.395119 &  2.197 \\
6.2 & 157.309956 & 26.392452 & \\
6.3 & 157.308576 & 26.388813 & \\
 \hline
 \end{tabular}
\flushleft{$^a$ See Figures~\ref{fig:sdss1029rgb_all} and
  \ref{fig:zoom_all} for the configuration of the multiple image
  systems and the morphology of each image. }
\flushleft{$^b$ Redshift ranges in brackets are $2\sigma$ redshift
  ranges predicted by the strong lens models.} 
\end{table}

We parametrically model the lens using the public software
{\it glafic} \citep{oguri10b}, although in the iterative process to
identify the multiple images we partly used the public software 
{\it lenstool} \citep{jullo07} as well. The mass model mainly
consists of dark halo components modelled by elliptical
\citet[][hereafter NFW]{navarro97} profiles with the radial profile of
$\rho(r)\propto r^{-1}(r+r_s)^{-2}$ and member galaxies
modelled as elliptical pseudo-Jaffe models
\citep[e.g.,][]{cohn01} with the radial profile of 
$\rho(r)\propto r^{-2}(r^2+r_{\rm cut}^2)^{-1}$. Since the cluster
core contains two bright galaxy concentrations, we include two
cluster-scale dark halo components centred at galaxy G1a
(R.A.=157.305789, Decl.=26.392602) 
and G2 (R.A.=157.302083, Decl.=26.392344). Each dark halo component
has 4 parameters: the ellipticity and position angle of their
isodensity contours, the virial mass $M_{\rm vir}$, and the
concentration parameter $c_{\rm vir}$. To reduce the number of
parameters, we assume that the velocity dispersion $\sigma$ and the
cutoff radius $r_{\rm cut}$ of the pseudo-Jaffe models scale with
the luminosities of the galaxies as $\sigma\propto L^{1/4}$ and
$r_{\rm cut}\propto L^{1/2}$, and regard the normalisations of the
scaling relations as free parameters. However, we do not apply these
scaling relations for one member galaxy just North of images 3.4--3.6
because the locations of these images are very sensitive to the
properties of this galaxy. The velocity dispersion and 
cutoff radius of this particular galaxy are included as free 
parameters. The ellipticity and position angle of the pseudo-Jaffe
model for each member galaxy are fixed to the value measured from the
{\it HST} ACS/F814W image. In addition to the dark halo and 
member galaxy components, we include four perturbation terms (lens
potential $\phi\propto r^2\cos m\theta$) with $m=2$ (external shear),
$3$, $4$, and $5$ which effectively describe the asymmetries of
cluster mass distributions that are commonly seen in simulations
\citep[e.g.,][]{meneghetti07}. Since we have no spectroscopic redshift
measurements for the multiply imaged galaxies, the redshifts of the
lensed galaxies are also treated as the free parameters. 

We use the positions of the 27 multiple images summarized in
Table~\ref{tab:img} as observational constraints. We assume positional
errors of $0\farcs1$ for the quasar images and $0\farcs4$ for the galaxy
images. These errors are larger than the measurement uncertainties, and
are chosen to take account of any complexity of the lens potential that
is not accounted for in the parametric mass model. For the
quasar images, we include flux ratios of $A/B=0.802$ and $C/B=0.455$
measured in the {\it Very Large Array} (VLA) radio image
\citep{kratzer11}, but assuming a conservative 20\% error on the flux
ratios. The assumed error is larger than the measurement errors 
(peak flux densities of 0.206, 0.257, 0.117~mJy/beam for images A, B,
and C, with an error of 0.013~mJy/beam for each component) in order to
take account of time variabilities as well as the complexity of the
lens potential. Recently, we measured time delay between quasar images
A and B 
as $\Delta t_{AB} = 744\pm10$~days \citep{fohlmeister12}. We include
this measurement as an observational constraint, although we assume a
5\% error on $\Delta t_{AB}$, $\sigma(\Delta t_{AB})=37$~days, which
is larger than the measurement error, because the time delay depends
also on the assumed Hubble constant and is limited by cosmic variance
due to matter fluctuations along the line-of-sight
\citep[e.g.,][]{barkana96}. In total, we have 57 observational 
constraints and 39 model parameters (8 from the dark halo components,
4 from the member galaxy components, 8 external perturbations, 5
source redshifts, and 14 source positions). Thus, the models have 18
degrees of freedom.

\subsection{The best-fit model}

\begin{figure}
\begin{center}
 \includegraphics[width=0.95\hsize]{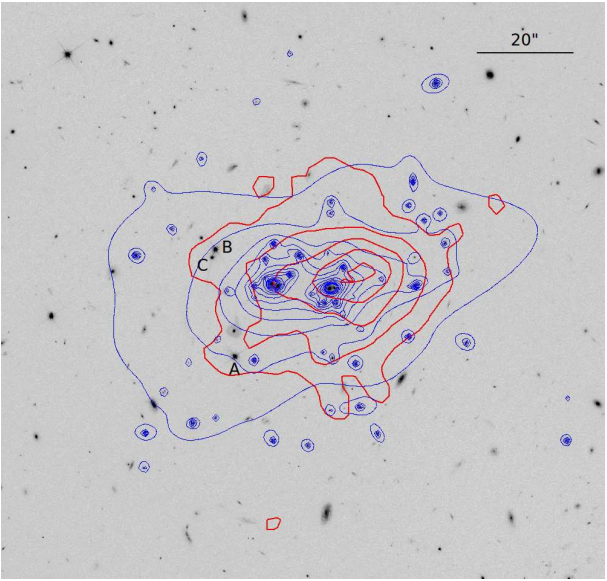}
\end{center}
\caption{Contours of the best-fit surface mass density profile from
  strong lens modelling ({\it thin blue}) and the {\it Chandra} X-ray surface
  brightness profile ({\it thick red}) are overlaid on the {\it HST} ACS/F814W
  image. In the X-ray contours we removed the quasar images, but there
  are some subtraction residuals that distort the X-ray contours. 
  \label{fig:sdss1029i_slx}}
\end{figure}

We derive the best-fit mass model by searching for the minimum
$\chi^2$. To speed-up the calculation, we estimate $\chi^2$ in the
source plane, which has been shown to be reasonably accurate for most 
of strong lens modelling purposes \citep[see, e.g.,][]{oguri10b}.
We find a $\chi^2$ for the best-fit model of $\chi^2=30$, which
is significantly larger than the number of degrees of
freedom $N_{\rm dof}=18$. In fact, a large fraction of the $\chi^2$
comes from the constraints on the flux ratio. While our input flux
ratios are $A/B=0.802$ and $C/B=0.455$, the best-fit model predicts
$A/B=0.082$ and $C/B=1.14$. This results in a large $\chi^2$
contribution despite our conservative errors on the flux ratios, and
indicates that the flux ratio anomaly found in previous simple mass
models \citep{oguri08,kratzer11} persists even in this much more
complex mass model. Once the flux ratios are removed from the
observational constraints, the best-fit model yields $\chi^2=10$ for
$N_{\rm  dof}=18-2=16$ degrees of freedom, suggesting that our choice
of the positional errors on the image locations is reasonable. In
particular, our best-fit model has a time delay of  $\Delta t_{AB}
=737$~days that reproduces the observed AB time delay almost
perfectly. We explore the flux ratio anomaly further in
Section~\ref{sec:flux}. 

The critical curves of the best-fit model are displayed in
Figure~\ref{fig:sdss1029rgb_all}. We find that contributions from 
both dark halo components to the critical curves are significant, and
as a consequence, the centroid of the main critical curve responsible
for the three quasar images falls between G1a and G2. In the best-fit
model, the dark halo at G2 has a smaller virial mass but a larger
concentration than the dark halo at G1a (see
Table~\ref{tab:model}). As a result, the halo of G2 
contributes more mass inside the Einstein radius than the halo at
G1a. Because the mass and concentration parameters constrained from
strong lens modelling are quite degenerate with each other, for a
more robust comparison, we fix the 
concentration parameters to  $c_{\rm vir}=5$ and find best-fit masses
of $M_{\rm vir}\simeq 1.5\times 10^{14}h^{-1}M_\odot$ for the dark
halo at G2 and $M_{\rm   vir}\simeq 1.0\times 10^{14}h^{-1}M_\odot$ 
for the dark halo at G1a. Thus, the dark halo associated with G2 is
more massive, although the effect of the dark halo at G1a is very
significant. We compare the surface mass density of the best-fit model
with the {\it Chandra} brightness profile \citep{ota12} in
Figure~\ref{fig:sdss1029i_slx}. As pointed out in \citet{ota12}, the
X-ray surface brightness profile of the main cluster component is
centred at G2, with an additional subpeak North-West of the main peak,
which appears to be different from the structure of the mass
distribution found in the strong lens models.  On the other hand, the
position angles of the best-fit mass model, $\theta_e=106.3$~deg
(measured East of North) for the dark halo at G2 and $105.7$~deg for
the dark halo at G1a, are in good agreement with the position
angle of the X-ray surface brightness profile,
$\theta_X=110.6\pm5.2$~deg \citep{ota12}. 

An important quantity to characterise strong lens systems is the
Einstein radius $\theta_{\rm E}$, although the definition of the
Einstein radius is not unique, for such a complex lens
system. In this paper, we define the Einstein radius as the radius
such that the mean enclosed surface density equals the critical
surface density, 
\begin{equation}
\bar{\kappa}(<\theta_{\rm
  E})\equiv\frac{1}{\pi\theta_{\rm
    E}^2}\int_{|\bm{\theta}|<\theta_{\rm E}} 
\kappa(\bm{\theta})d\bm{\theta}=1,
\end{equation}
adopting the average of the locations of G2 and G1a (R.A.=157.303936,
Decl.=26.392473) as the centre. The Einstein radius of the best-fit
model is $\theta_{\rm E}=15\farcs2$ for the quasar redshift of
$z_s=2.197$, which is consistent with the extent of the main critical
curve shown in Figure~\ref{fig:sdss1029rgb_all}. However, we note that
the choice of the centre is somewhat arbitrary. If we move the centre
for calculating the Einstein radius by $\pm1''$ along the North-South
and East-West directions, the value changes in the range
$14\farcs8<\theta_{\rm E}<15\farcs4$.   

\begin{table*}
 \caption{Best-fit values of key model parameters from the strong lens
   models. The position angle $\theta_e$ is measured East of North.
  Aside from the galaxy near image 3.4-3.6, the parameters of the
  other member galaxies are scaled from those for G2 as $\sigma\propto
  L^{1/4}$ for the velocity dispersion and $r_{\rm cut}\propto
  L^{1/2}$ for the cutoff radius. The uncertainties are $2\sigma$
  confidence limits from the Markov Chain Monte Carlo analysis.
\label{tab:model}}    
 \begin{tabular}{@{}ccccccc}
 \hline
   component
   & $M_{\rm vir}$
   & $c_{\rm vir}$
   & $\sigma$
   & $r_{\rm cut}$
   & $e$
   & $\theta_e$
   \\
   & ($10^{14}h^{-1}M_\odot$) 
   & 
   & (${\rm km\,s^{-1}}$)
   & (arcsec)
   & 
   & (deg)
 \\
 \hline
dark halo at G1 &  $1.14^{+5.35}_{-2.20}$ & $5.49^{+3.91}_{-3.53}$ 
  & $\cdots$ & $\cdots$
  & $0.74^{+0.07}_{-0.15}$ & $106.3^{+1.4}_{-3.8}$ \\
dark halo at G2 &  $3.04^{+1.79}_{-2.76}$ & $2.94^{+7.29}_{-0.78}$ 
  & $\cdots$ & $\cdots$
  & $0.41^{+0.12}_{-0.33}$ & $105.7^{+3.8}_{-3.0}$ \\
galaxy G2 & $\cdots$ & $\cdots$
   & $291^{+98}_{-64}$ & $13.4^{+5.7}_{-11.2}$
   & ($\equiv 0.252$) & ($\equiv -57.6$)\\
galaxy near 3.4-3.6 & $\cdots$ & $\cdots$
   & $141^{+70}_{-38}$ & $>1.4$
   & ($\equiv 0.313$) & ($\equiv -167.9$)\\
 \hline
 \end{tabular}
\end{table*}

We explore the errors on the model parameters employing a Markov
Chain Monte Carlo technique. We adopt Metropolis-Hastings sampling
with a multivariate-Gaussian for the proposal distribution.  From this 
procedure we estimate the possible range of redshifts for the multiply
imaged galaxies, which we summarize in Table~\ref{tab:img}. 
We also estimate errors on key model parameters of the best-fit strong
lens mass model, which are summarized in Table~\ref{tab:model}.
For comparison, we also derive photometric redshifts of the multiply
imaged galaxies based on the three-band {\it HST} photometry using the
public code EAZY \citep{brammer08}, and find that the photometric
redshifts are in good agreement with the model predictions for ID 1,
2, and 5. The photometric redshifts are lower than the model
predictions for ID 2 and 3, although the upper bounds on
their photometric redshifts of $z_p\la 2.6$ ($2\sigma$) marginally
overlap with the expected redshift ranges from strong lens modelling. 
In addition, we find that our strong lens modelling constrains the
Einstein radius to the $2\sigma$ range of $14\farcs5<\theta_{\rm
  E}<15\farcs5$. Considering the additional uncertainties on the
cluster centre mentioned above, we adopt $\theta_{\rm E}=15\farcs2\pm
0\farcs5$ ($1\sigma$ error) for the combined lensing analysis in
Section~\ref{sec:comb}.  

\section{Weak lensing analysis}
\label{sec:wlens}

\subsection{HST weak lensing measurement}

We carry out the weak lensing analysis using the {\it HST} ACS/F814W
image. We closely follow \citet{schrabback10} for the shear
measurement procedure, which is based on the object catalogue
generated by SExtractor \citep{bertin96} and the weak lensing method
outlined in \cite{kaiser95}. In particular, we model temporal point spread
function (PSF) variations based on a principal component analysis of
dense stellar fields. To correct for the systematic shear
underestimates of the KSB method due to noise bias, we include an
empirical calibration bias as a function of the signal-to-noise ratio
of each galaxy. The method was tested against {\it HST} ACS-like
simulation data and was shown to be accurate to better than 2\% over
the entire magnitude range used \citep[see][for details]{schrabback10}. 

A possible systematic effect in cluster weak lensing analysis is a
dilution of the weak lensing signals by member galaxies 
\citep[e.g.,][]{broadhurst05,medezinski07,medezinski10}.
In addition to member galaxies in the red-sequence (see
Figure~\ref{fig:colmag}), the galaxy catalogue may contain cluster 
member galaxies up to $\sim 0.5$~mag bluer than the red sequence 
\citep{medezinski07}. Hence, in order to reduce the dilution, we
apply a colour cut of 
\begin{equation}
m_{\rm F475W}-m_{\rm F814W}<-0.15(m_{\rm F814W}-22)+2.2,
\label{eq:ccut}
\end{equation}
for our weak lensing analysis, where the magnitudes are the $1''$
diameter aperture magnitudes shown in Figure~\ref{fig:colmag}. In
addition, we apply a total magnitude cut of $m_{\rm F814W}<27$ where
the number counts of the galaxies in the data turns  over. The final
source galaxy number density of $\sim 75$~arcmin$^{-2}$ after applying
these selection limits is still large enough to allow high-resolution
weak lensing measurements near the cluster core. 

We estimate the mean depth $\langle D_{ls}/D_{os}\rangle$ of our
source galaxy sample using the redshift distribution of galaxies as
a function of $m_{F814W}$ derived by \citet{schrabback10}. Although
the colour cut (Equation~\ref{eq:ccut}) can in principle modify the
redshift distribution, we here ignore this effect because a majority
of galaxies outside the red-sequence are not removed by this colour cut
(see Figure~\ref{fig:colmag}) and therefore the effect of the colour 
cut on the redshift distribution is expected to be small. As a cross
check, we also estimate the mean depth using a photometric redshift
catalogue in the Hubble Ultra Deep Field \citep{coe06}. After applying
the same colour (Equation~\ref{eq:ccut}) and magnitude cuts, we obtain
a mean depth that is quite consistent with the earlier estimate. We then
estimate the effective source redshift at which the distance ratio
becomes equal to the mean depth. We find an effective source redshift
of  $z_{s,{\rm eff}}=1.06$, which we adopt for our {\it HST} weak
lensing analysis throughout the paper.  

\subsection{Result}

\begin{figure}
\begin{center}
 \includegraphics[width=0.95\hsize]{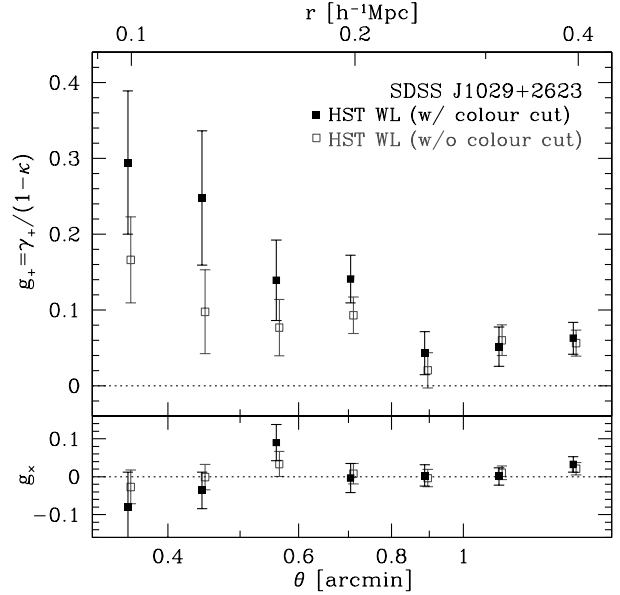}
\end{center}
\caption{Tangential shear profiles from the weak lensing analysis of the
  {\it HST} ACS/F814W image. Filled squares show the shear profile
  after the colour cut to remove cluster member galaxies
  (equation~\ref{eq:ccut}), while open squares are the shear profile
  without the colour cut. The lower panel shows the radial profile of
  the $45^\circ$ rotated component.
\label{fig:profile_wl}}
\end{figure}

Figure~\ref{fig:profile_wl} shows the tangential shear profile from
the {\it HST} ACS/F814W source galaxy catalogue. Here we again adopt
the average of the locations of G2 and G1a as the cluster centre,
although we note that the tangential shear profile is not very
different even if we adopt the location of G1a or 
G2 as a cluster centre. The tangential shear detections are
statistically significant, with increasing shear toward the cluster
centre. To check the impact of dilution by cluster member galaxies, we
derived a tangential shear profile from the source galaxy catalogue without
the colour cut defined in equation~(\ref{eq:ccut}). The resulting
shear profile, which is also shown in Figure~\ref{fig:profile_wl},
has a significantly weaker amplitude compared with the profile
including the colour cut, and the effect is stronger near the cluster
centre. This confirms earlier claims that proper selection of galaxies
to remove cluster members is essential for accurate weak lensing
analysis. We also check the radial number density profile of the
source galaxies and find that the number counts after the colour 
cut decreases toward the cluster centre, as expected from the lensing
magnification effect \citep[e.g.,][]{umetsu08} and masking by
cluster member galaxies. 

\begin{figure}
\begin{center}
 \includegraphics[width=0.99\hsize]{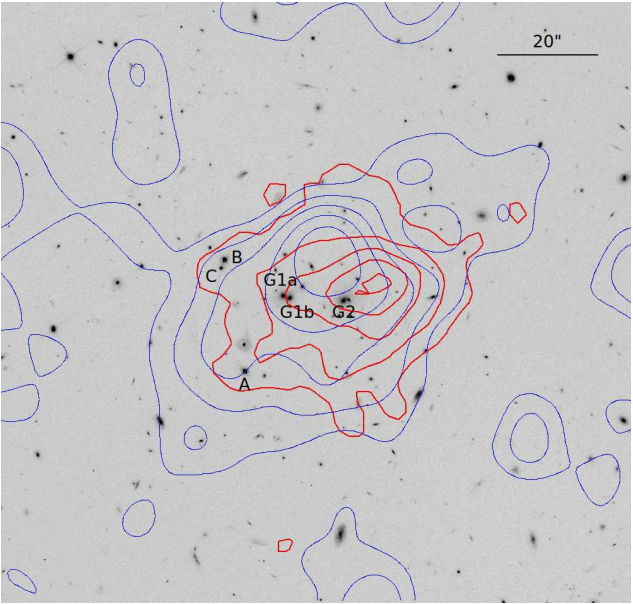}
\end{center}
\caption{Similar to Figure~\ref{fig:sdss1029i_slx}, but comparing the
  surface mass density profile from the weak lensing analysis of the
  {\it HST} ACS/F814W image ({\it thin blue}) to the {\it  Chandra} X-ray 
  surface brightness profile ({\it thick red}). The mass map is
  Gaussian-smoothed with $\sigma=8''$. 
\label{fig:sdss1029i_wlx}}
\end{figure}

Next we explore the two-dimensional mass map reconstructed from weak
lensing, using the inversion technique of \citet{kaiser93}. As shown
in Figure~\ref{fig:sdss1029i_wlx}, there is a significant weak 
lensing mass peak around galaxy G1a and G2, which is consistent with
the strong lens modelling result shown in Section~\ref{sec:slens}.
For more quantitative comparisons with the {\it Chandra} X-ray surface
brightness profile, we fit the binned, unsmoothed two-dimensional
shear map with an elliptical NFW profile, following the methodology
developed in \citet{oguri10a} and \citet{oguri12}. The model involves
6 free parameters: the virial mass and concentration parameter, the
ellipticity and position angle, and the centroid. We find that the
{\it HST} weak lensing data constrain the centroid of the surface mass
density profile to ($\Delta x$, $\Delta y$)=($1.2^{+6.2}_{-7.4}$, 
$1.7^{+6.7}_{-8.9}$), in arcseconds from the average of the locations
of G1a and G2, where positive directions of $\Delta x$ and $\Delta y$
correspond to West and North, respectively.   
Note that ($\Delta x$, $\Delta y$)=($\mp5.975$, $\pm0.465$) correspond
to the locations of G1a and G2. Thus, the two-dimensional weak lensing
analysis indicates that the mass centroid is quite consistent with 
the average location of G1a and G2, but it is also consistent with 
the location of G1a or G2. In addition, we find that the shear map
prefers an elongated mass distribution, with ellipticity
$e=0.72^{+0.13}_{-0.27}$ and position angle
$\theta_e=93^{+15}_{-11}$~deg. The position angle is in good agreement
with both the position angle from strong lens modelling and that of
the X-ray surface brightness profile. 

\section{Combined lensing analysis}
\label{sec:comb}

\subsection{Radial density profile}
\label{sec:raden}

\begin{figure}
\begin{center}
 \includegraphics[width=0.99\hsize]{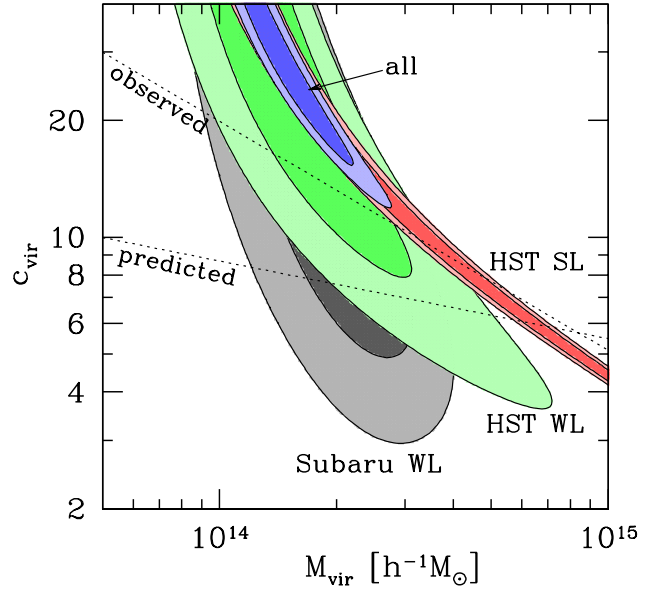}
\end{center}
\caption{Constraints on the virial mass $M_{\rm vir}$ and the
  concentration parameter $c_{\rm vir}$ from the {\it HST} strong
  lensing (SL) analysis, {\it HST} weak lensing (WL), and Subaru weak
  lensing. Here the strong lensing constraint comes from the
  constraint on the Einstein radius, $\theta_{\rm E}=15\farcs2\pm0\farcs5$ 
   for $z_s=2.197$. The weak lensing constraints are derived from
  azimuthally averaged tangential shear profiles (see
  Figure~\ref{fig:profile_all}). 
  For each constraint, we show $1\sigma$ and $2\sigma$ ranges in this
  parameter space. The innermost contours show the combined
  constraints. Dotted lines indicate theoretically predicted and
  observed average mass-concentration relations for strong lensing
  clusters at $z\sim 0.5$ derived in \citet{oguri12}. 
  \label{fig:cont_mc}}
\end{figure}
 
\begin{table}
 \caption{Best-fit mass and concentration parameters from combined
   lensing analysis (see also Figure~\ref{fig:cont_mc}). Errors
   correspond to $1\sigma$ confidence level. 
\label{tab:mc}}    
 \begin{tabular}{@{}ccc}
 \hline
   constraints
   &  $M_{\rm vir}$
   &  $c_{\rm vir}$
   \\
   & ($10^{14}h^{-1}M_\odot$) 
   & \\
 \hline
{\it HST}-WL &  $1.35^{+0.89}_{-0.28}$ & $30.9^{+8.9}_{-19.2}$ \\
Subaru-WL    &  $2.00^{+0.70}_{-0.58}$ & $11.5^{+14.2}_{-5.0}$\\
{\it HST}+Subaru-WL &  $1.74^{+0.55}_{-0.45}$ & $17.0^{+19.3}_{-6.5}$\\
{\it HST}-WL+SL & $1.32^{+0.63}_{-0.14}$ & $34.7^{+5.1}_{-16.5}$\\ 
{\it HST}+Subaru-WL+SL &  $1.55^{+0.40}_{-0.35}$ & $25.7^{+14.1}_{-7.5}$\\
 \hline
 \end{tabular}
\end{table}

We now combine the constraints from strong lensing
(Section~\ref{sec:slens}), {\it HST} weak lensing
(Section~\ref{sec:wlens}), and  Subaru weak lensing \citep{oguri12}
to derive constraints on the overall radial density profile of the
lensing cluster. For the strong lensing result, we conservatively use
the constraint on the Einstein radius of $\theta_{\rm
  E}=15\farcs2\pm0\farcs5$  for the source redshift of
$z_s=2.197$. For the {\it HST} and Subaru weak lensing data, we use
the tangential shear profiles as constraints. We use these 
data to constrain the virial mass $M_{\rm vir}$ and the concentration 
parameter $c_{\rm vir}$, assuming the NFW profile. The fit was
performed for the mass range of 
$10^{13}h^{-1}M_\odot<M_{\rm vir}<10^{16}h^{-1}M_\odot$ and
the concentration parameter range of $0.01<c_{\rm vir}<39.8$.

\begin{figure}
\begin{center}
 \includegraphics[width=0.99\hsize]{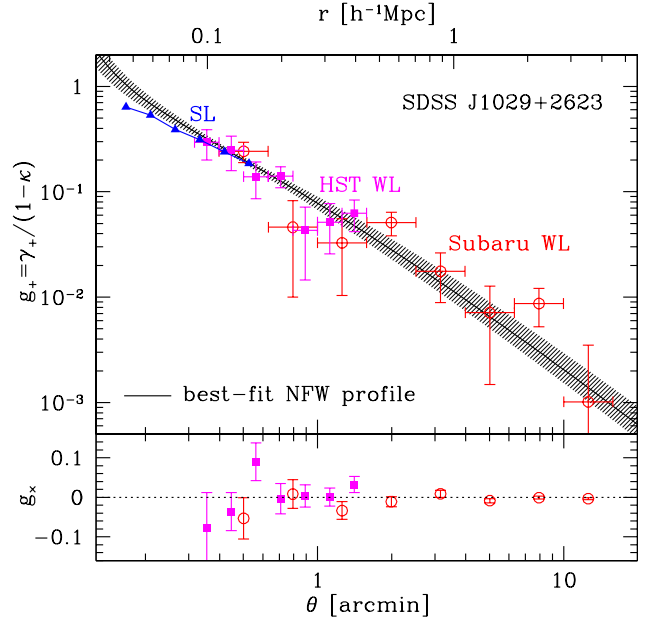}
\end{center}
\caption{The tangential shear profile of the best-fit NFW model ({\it
    solid line}) as compared to the {\it HST} ({\it filled squares}) and
  Subaru ({\it open circles}) shear profiles and the reduced
  tangential shear $g_+$ predicted by the best-fit strong lens model ({\it
    filled triangles}) for $z_s=1.1$. The best-fit shear profile is
   computed for $z_s=1.1$, which is close to the effective source
  redshifts of $z_s=1.06$ and $1.19$ for the {\it HST} and Subaru weak
  lensing data. The shaded region indicates the $1\sigma$ range. 
  The lower panel shows the radial profile of the $45^\circ$
  rotated component $g_\times$. 
\label{fig:profile_all}}
\end{figure}

We show constraints in the $M_{\rm vir}$-$c_{\rm vir}$ plane in
Figure~\ref{fig:cont_mc}, and summarize the resulting constraints on
$M_{\rm vir}$ and $c_{\rm vir}$ in Table~\ref{tab:mc}. 
The individual constraints are quite
degenerate in this plane, but along slightly different directions so
that we obtain a tighter constraint when combining the three
constraints. We find $M_{\rm 
vir}=1.55^{+0.40}_{-0.35}\times 10^{14}h^{-1}M_\odot$ and 
$c_{\rm vir}=25.7^{+14.1}_{-7.5}$ from the 
combined analysis. The large concentration parameter value for the
virial mass of $\sim 10^{14}h^{-1}M_\odot$ is in line with a recent
study the $M_{\rm vir}$-$c_{\rm vir}$ relation for strong lensing
clusters \citep{oguri12}, which can be explained by the effect of
baryon cooling and central galaxies \citep[][but see also
\citealt{duffy10,mccarthy10}]{fedeli12}. However, the 
high concentration for this cluster should be interpreted cautiously
given the complex core structure with two density peaks. 
We compare the tangential shear profiles with the best-fit model
prediction in Figure~\ref{fig:profile_all}. The three observational
constraints are complementary and they are consistent with each other
where they overlap. The best-fit NFW profile reproduces the
observations for a wide range in radii.

Here we discuss several systematic effects that can affect
our results. One is the redshift distribution of source galaxies for
weak lensing analysis. Since the area of the field used for our weak
lensing analysis is small, the redshift distribution in this field can
be different from that of a mean distribution. Another effect is a
multiplicative shear measurement bias, which is estimated to be $\la
5$\% for our Subaru weak lensing measurements. Also the {\it HST} weak
lensing measurement probes the region $g\ga 0.1$, where weak lensing
measurements are less tested and therefore less reliable. To estimate
a possible impact of these systematic errors, we consider an extreme
situation where both the {\it HST} and Subaru weak lensing
measurements are offset by $\pm 10$\%, and find the resulting shifts
of the best-fit virial mass to be
$\sim \pm 0.3\times 10^{14}h^{-1}M_\odot$. The systematic error for
this case is still comparable to the $1\sigma$ statistical error,
implying that these systematic errors are not significant.  
 
\subsection{Comparison with X-ray mass}

\begin{figure}
\begin{center}
 \includegraphics[width=0.99\hsize]{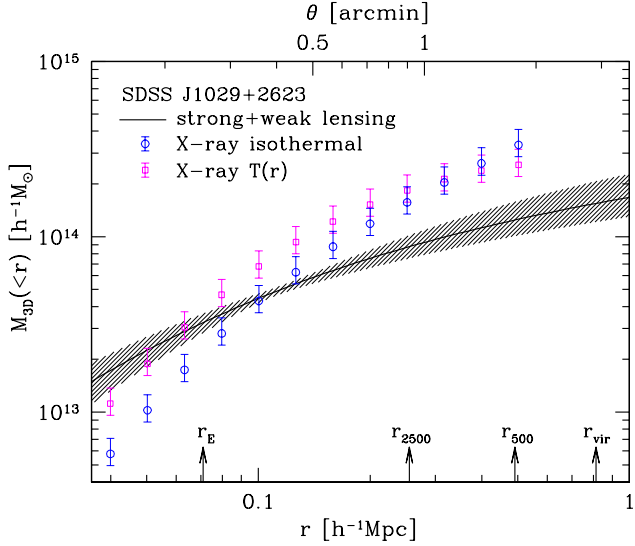}
\end{center}
\caption{The enclosed mass within a sphere of radius $r$ from the
  {\it Chandra} X-ray analysis and from the combined lensing analysis
  ({\it solid line}; see Section~\ref{sec:raden}), assuming a
  spherically symmetric mass distribution. For the {\it Chandra} X-ray
  analysis, we show the results for the isothermal $\beta$-model
  from \citet[][{\it open circle}]{ota12} and a $\beta$-model
  with the temperature profile of \citet[][{\it open
      squares}]{burns10}. The errors in the X-ray masses 
  come from the statistical errors in the X-ray temperature
  measurement. For reference, we also show the cluster radii 
  $r_{\rm vir}$, $r_{500}$, and $r_{2500}$, which are computed from
  the best-fit lensing mass profile, as well as the Einstein radius
  $r_{\rm E}$. 
\label{fig:m3d}}
\end{figure}

\citet{ota12} presented the {\it Chandra} X-ray analysis of
SDSS~J1029+2623, and derived a mass profile assuming hydrostatic
equilibrium and isothermality. Figure~\ref{fig:m3d} compares the X-ray
mass profile from \citet{ota12} with the result of the combined
lensing analysis in 
Section~\ref{sec:raden}. We find that the mass profiles derived from
lensing and X-ray differ significantly with each other. While the
enclosed masses agree at the radius of $\sim 100 h^{-1}{\rm kpc}$
that roughly corresponds to the Einstein radius of this system, the
enclosed masses inferred from the X-ray data are a factor of $\sim 2$
larger than those inferred from the combined lensing analysis at radii
$r\ga r_{2500}$, where $r_{2500}$ is defined by the radius within
which the average density is 2500 times the critical density at the
cluster redshift. We note that the lensing derived mass implies an
X-ray temperature and luminosity of $T\sim 2-3$~keV and 
$L_X\sim 10^{44}$~erg~s$^{-1}$ based on X-ray 
scaling relations \citep[e.g.,][]{dai07}, which are significantly
lower than the observed X-ray temperature of $T\sim 8.1$~keV and $L_X\sim
10^{45}$~erg~s$^{-1}$ \citep{ota12}. It is also worth noting that
recent systematic weak lensing analysis of clusters at $z\la 1$ found
somewhat smaller weak lensing masses for a given X-ray temperature
\citep{jee11}, which is qualitatively similar to our result, although
the difference appears to be much smaller than found here.

There are several effects that can lead to a systematic difference
between the X-ray and lensing masses. One of the most significant
effects for X-ray cluster mass measurements is any violation of the
assumption of hydrostatic equilibrium due the presence of non-thermal
pressure support. However, this effect typically leads to an
underestimate of the X-ray mass, particularly in the outskirts of
clusters \citep[e.g.,][]{mahdavi08,zhang10}, and would only make the
discrepancy larger. Another possibility is our assumption of
isothermality. In fact, X-ray temperatures generally decrease at large
radii, but our {\it Chandra} data were not sensitive enough to detect
such a decrease. To examine this effect, we adopt a temperature 
profile obtained from hydrodynamic simulations \citep{burns10}, which
appears to be consistent with recent X-ray observations of cluster
outskirts \citep[e.g.,][]{akamatsu11}, and recalculate the X-ray mass
profile. While this reduces the difference at the center, it cannot
explain the overall difference between the X-ray and lensing 
masses. It is also difficult to explain the difference by a
triaxial halo, because the large $c_{\rm vir}$ implies that the
major-axis of the cluster is more likely to be aligned with the
line-of-sight direction, in which case the lensing mass should be
overestimated \citep{oguri05}. 

The most likely explanation for the mass discrepancy is shock heating
of the intracluster gas during a merger. Numerical simulations show
significant boosts of X-ray luminosity and temperature $\sim 1$~Gyr
after mergers, which can lead to significant overestimates of X-ray
masses \citep[e.g.,][]{ricker01,takizawa10,rasia11,nelson12}.
Observationally, there are several clusters showing signs of ongoing
mergers that also have significantly higher X-ray masses than lensing
masses \citep[e.g.,][]{okabe08,okabe11,soucail12}. Indeed
the lensing cluster of SDSS~J0129+2623 shows hints of an ongoing
merger, including the bimodal nature of cluster cores, the complex
X-ray morphology, a possible offset between mass and X-ray
centroids, and the large G1-G2 velocity difference. In addition, a
line-of-sight merger can naturally explain the high concentration
estimates for this cluster \citep[e.g.,][]{king07}.
Spectroscopy of many more cluster member galaxies is needed to
understand this complex cluster further. If this interpretation is
correct, the agreement between X-ray and lensing masses near the
Einstein radius might just be a coincidence, in that both the masses
are overestimated by merger shock heating and halo elongation
along the line-of-sight, respectively.   

\subsection{Gas-to-mass ratio}

Another useful cross-check of our interpretation is provided by the
gas-to-mass ratio, $f_{\rm gas}(<r)=M_{\rm gas}(<r)/M_{\rm tot}(<r)$,
because it is expected to roughly match the cosmic baryon fraction
$\Omega_b/\Omega_M\simeq0.167$ \citep{komatsu11} for massive clusters,
and also because $M_{\rm gas}$ appears to be the most promising
cluster mass proxy \citep{okabe10,fabjan11}. We use the gas mass
profile implied by the isothermal $\beta$-model from \citet{ota12}
to estimate a gas mass within the radius $r_{500}=0.49h^{-1}$Mpc
from the combined lensing analysis. We find a gas mass fraction of
$f_{\rm gas}(<r_{500})=0.060\pm0.014$, which is significantly smaller
than the cosmic baryon fraction. The gas mass fraction is slightly
increased to $f_{\rm gas}(<r_{500})=0.077\pm0.017$ when the
temperature profile of \citet{burns10} is used to derive the
total mass, but the value is still low. On the other hand, when the
lensing mass estimate is used for the total mass, we obtain a gas
mass fraction of  $f_{\rm gas}(<r_{500})=0.158\pm0.038$, which is in
better agreement with $f_{\rm gas}$ estimates in  other clusters
\citep[e.g.,][]{vikhlinin06,allen08,ettori09,umetsu09,dai10b,okabe10,pratt10}.
This further supports our interpretation that the X-ray derived mass
is significantly overestimated due to shock heating of the
intracluster gas during a merger.

\section{Implications for the flux ratio anomaly}
\label{sec:flux}

\subsection{Searching for substructures}

\begin{figure}
\begin{center}
 \includegraphics[width=0.85\hsize]{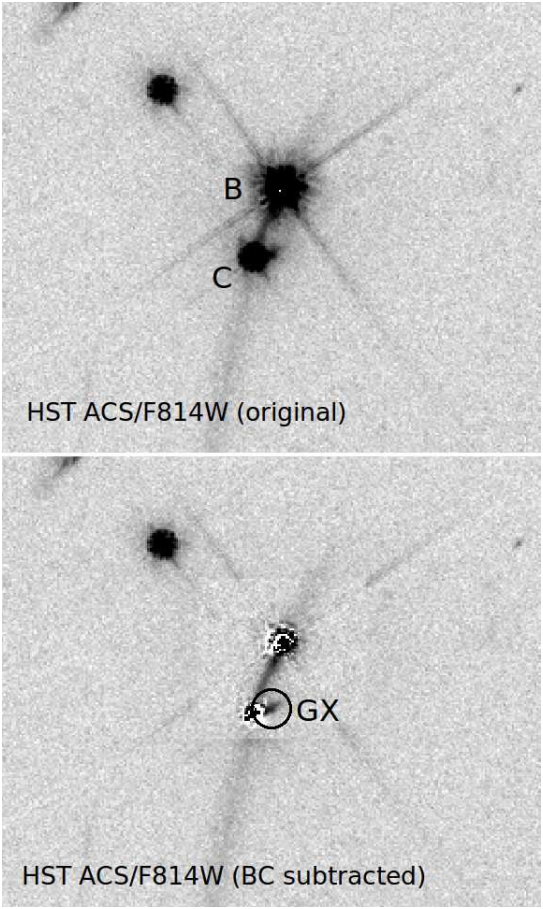}
\end{center}
\caption{{\it Top:} The {\it HST} ACS/F814W image around the merging
  quasar images B and C. {\it Bottom:} The image after subtracting
  quasar images. The probable galaxy GX is marked by a circle. We used
  a nearby star as a PSF template.
\label{fig:sdss1029bc_sub}}
\end{figure}

Our strong lens models in Section~\ref{sec:slens} confirm the
earlier claims \citep{oguri08,kratzer11,ota12} that a substructure 
is required to explain the anomalous flux ratios between the three 
quasar images. In order to search for possible substructures, we
subtract the quasars from each {\it HST} image and carefully examine
the residual images. In the ACS/F814W image, we find a probable galaxy
in the vicinity of image C, as shown in
Figure~\ref{fig:sdss1029bc_sub}. We fit the galaxy, named GX, assuming
a Sersic profile with index $n=4$ using {\it galfit} \citep{peng02},
and find the centroid of the galaxy to be $\sim 0\farcs4$ from image C
(R.A.=157.309482, Decl.=26.394158), and that the magnitude of 
the galaxy is $m_{F814W}\sim 24.3$. These estimates have large
uncertainties due to the significant residuals from the  subtraction
of image C (see Figure~\ref{fig:sdss1029bc_sub}). In the strong lens
models we estimate that galaxy G2 with $m_{F814W}=19.62$ has a
velocity dispersion of $\sigma=291\,{\rm km\,s^{-1}}$, which implies a
velocity dispersion of $\sigma\simeq 100\,{\rm km\,s^{-1}}$ for GX 
assuming $\sigma\propto L^{1/4}$. The inferred velocity dispersion of
GX is large enough to mean that it will significantly perturb the flux
of image C. 

\subsection{Mass modelling}

\begin{figure}
\begin{center}
 \includegraphics[width=0.95\hsize]{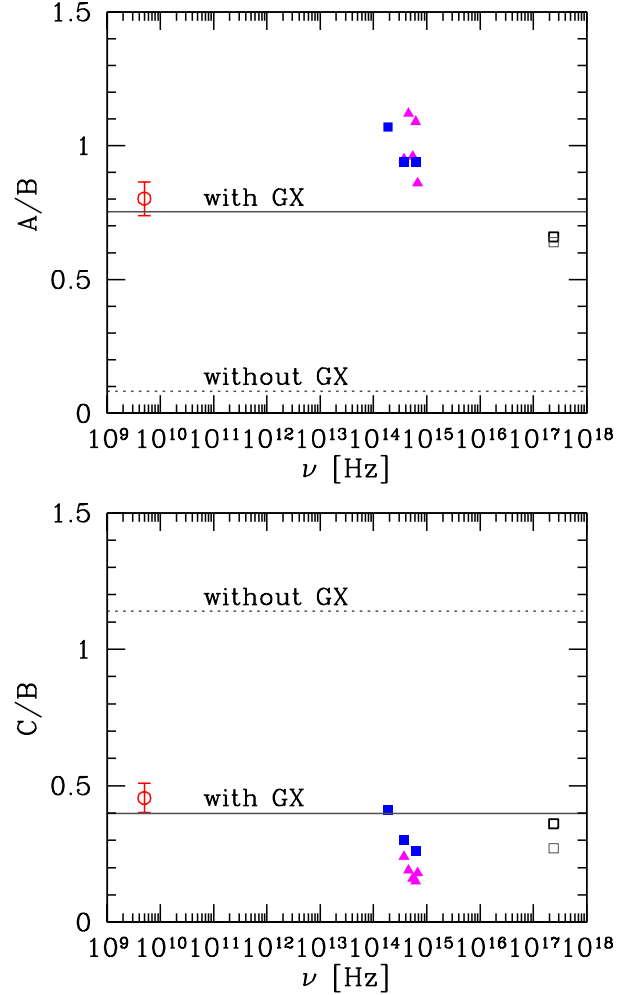}
\end{center}
\caption{Flux ratios of the best-fit strong lens models with ({\it solid})
  and without ({\it dotted}) galaxy GX are compared with the observed
  flux ratios in different wavelengths. The top panel shows the flux ratio
  between images A and B, while the bottom panel shows the flux
  ratio between images B and C. Observed flux ratios are obtained from
  the VLA radio image ({\it open circle}) by \citet{kratzer11}, the
  {\it HST} images ({\it filled squares}) presented in this paper,
  ground-based optical images ({\it filled triangles}) by
  \citet{oguri08}, and the {\it Chandra} X-ray image with ({\it black
    open square}) and without ({\it grey thin open square}) the
  absorption correction by \citet{ota12}.
\label{fig:fratio}}
\end{figure}

We next explore the impact of GX on the strong lens models. Here we
assume the same mass model as used in Section~\ref{sec:slens}, but
include an additional component modelled by pseudo-Jaffe model at the
location of GX with the velocity dispersion and the cutoff radius as
free parameters. We find that the best-fit model has $\chi^2=7.5$
including the flux ratios, which is significantly smaller than the
$\chi^2$ of the original best-fit model ($\chi^2=30$) and the large
improvement is due to  
improvements in fitting the flux ratios. This confirms the argument by
\citet{kratzer11} that a substructure in the vicinity of image C can
explain the anomalous flux ratio. The velocity dispersion of galaxy GX
in the best-fit model is $\sigma=76\,{\rm km\,s^{-1}}$, which is broadly
consistent with the velocity dispersion expected from the
magnitude. We summarize the flux ratios of the best-fit mass models,
as well as the observed flux ratios at various wavelengths, in
Figure~\ref{fig:fratio}.  

The best-fit models with and without galaxy GX have quite different
total magnification factors. For the model without galaxy GX, the
magnifications for the quasar images are $\mu_A=6.8$, $\mu_B=82.9$,
and $\mu_C=94.7$, whereas for the model with galaxy GX, they are
$\mu_A=10.4$, $\mu_B=13.8$, and $\mu_C=5.5$. The total magnifications
are $\mu_{\rm tot}=184.4$ for the model without GX and $\mu_{\rm
  tot}=29.7$ for the model with GX.  
   
Substructures near merging image pairs can have large effects on time
delays between the images \citep{oguri07,keeton09}. This was indeed
the case for SDSS~J1004+4112 \citep{inada03}, another large-separation
lensed quasar produced by a massive cluster of galaxies, in which 
a time delay between the smallest image pair was found to be affected
by the perturbation from nearby cluster member galaxies
\citep{fohlmeister07}. We find that our best-fit mass models
predict the time delay between image B and C as $\Delta
t_{BC}=0.7$~day for the model without GX, and $\Delta t_{BC}=7.7$~day
for the model with GX. In both cases, image B leads image C as
expected from the image geometry. On the
other hand, GX has a negligible impact on the AB time delay. The large
difference of $\Delta t_{BC}$ implies that measuring the BC time delay
will greatly help constrain the substructure that affects these quasar
images.  

\subsection{Morphology of the lensed host galaxy}

\begin{figure*}
\begin{center}
 \includegraphics[width=0.99\hsize]{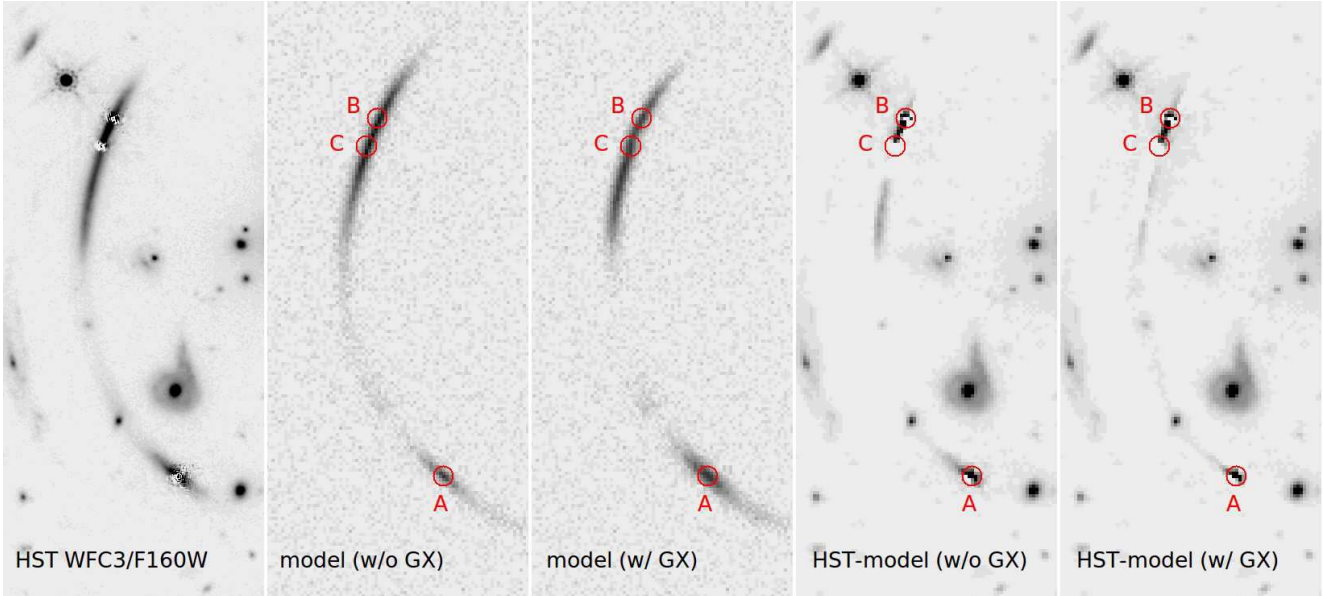}
\end{center}
\caption{{\it First panel:} The {\it HST} WFC3/F160W image showing the
  lensed host galaxy of the quasar at $z_s=2.197$. All the quasar
  components are subtracted using a nearby star as a PSF template. 
  {\it Second panel:} The morphology of the lensed host galaxy in the
  best-fit mass model without galaxy GX from 
  Section~\ref{sec:slens}. The locations of the three quasar images 
  are marked by circles. 
  {\it Third panel:} Similar to the middle panel, but the case for the
  best-fit model with galaxy GX.
  {\it Fourth panel:} The {\it HST} WFC3/F160W image after subtracting
  the best-fit host galaxy model without galaxy GX.
  {\it Fifth panel:} The {\it HST} WFC3/F160W image after subtracting
  the best-fit host galaxy model with galaxy GX. Note that all the
  panels are presented with the same grey-scale stretch.
\label{fig:hostcomp}}
\end{figure*}

An interesting feature of SDSS~J1029+2623 is the presence of a highly
elongated host galaxy, which is most prominent in the {\it HST}
WFC3/F160W image. Here we examine whether our best-fit models
reproduce the morphology of the lensed host galaxy. To do so, we
include the host galaxy modelled by a Sersic profile, and optimise
the parameters (the total flux, size, ellipticity and position angle)
of the Sersic profile assuming the best-fit mass models, both with and
without galaxy GX. We fix the centroid of the host galaxy in the
source plane to the best-fit source quasar position. The optimisation
is conducted on a pixel-by-pixel basis using a binned {\it HST}
WFC3/F160W image with a pixel scale of $0\farcs2$. The model-predicted
images are convolved with a PSF assuming a Moffat profile with a FWHM
of $0\farcs18$ and $\beta=2$, before comparing with the observed image. 

The result is shown in Figure~\ref{fig:hostcomp}. We find that our
best-fit models reproduce the morphology of the lensed host galaxy
quite well. The match is better for the model with galaxy GX, which
further supports its presence and the significant impact of galaxy GX
on this lens system. In particular, we note that the observed host
galaxy shows signs of a  kink near image C (see also
Figure~\ref{fig:sdss1029bc_sub}) which is reproduced in our best-fit 
model with galaxy GX. We estimate the total magnitude of the lensed
host galaxy as $m_{F160W}\sim 18.5$ and the lensing-corrected original
magnitude as $m_{F160W}\sim 22.4$. The best-fit effective radius and
Sersic index are $r_e=0\farcs27$ and $n=0.88$, respectively. The
presence of galaxy GX has a negligible impact on the total
magnification of the host galaxy. The highly magnified host galaxy
provides a very unique opportunity to study the relation between
supermassive black holes and host galaxies in a distant 
quasar \citep[e.g.,][]{peng06,ross09}.  

\section{Conclusion}
\label{sec:conclusion}

We have presented a detailed gravitational lensing analysis of
SDSS~J1029+2623, the largest-separation lensed quasar system known
(image separation $\theta=22\farcs5$, lens redshift $z_l=0.584$, and
source redshift $z_s=2.197$), based on new {\it HST} ACS and
WFC3 observations. With the help of the high-resolution {\it HST} images,
we have identified 6 new multiply imaged galaxies in addition to
the three-image lensed quasar. Based on the newly identified multiple
images, we construct an accurate mass model at the cluster core. 
We find dark haloes with roughly similar masses should be associated 
with both of the bright central galaxies. The strong lens models allow
us to estimate the Einstein radius of the overall lens as 
$\theta_{\rm E}=15\farcs2\pm0\farcs5$. Our best-fit strong
lens model successfully reproduces the observed time delay between
quasar image A and B which was recently measured by 
\citet{fohlmeister12}. We have also detected a significant weak
lensing signal in the {\it HST} ACS image. The mass map obtained from
the strong and weak lensing analysis appears to be different from the
X-ray surface brightness profile \citep{ota12}, suggesting an ongoing
merger.   

We have constrained the radial mass profile of the cluster from the 
{\it HST} strong and weak lensing analysis as well as the Subaru weak
lensing result \citep{oguri12}. The combination of the three lensing
constraints enabled a robust determination of the radial mass profile,
with a best-fit virial mass of $M_{\rm vir}=1.55^{+0.40}_{-0.35}\times  
10^{14}h^{-1}M_\odot$ and a concentration parameter of 
$c_{\rm vir}=25.7^{+14.1}_{-7.5}$. The lensing derived
mass profile differs significantly from the mass profile inferred from
the {\it Chandra} X-ray data assuming hydrostatic equilibrium and
isothermality, such that the X-ray mass is a factor of $\sim 2$
larger at outer radii of $r_{2500}\la r \la r_{500}$. We ascribe the
mass discrepancy to shock heating of the intracluster gas during a
merger, which leads to significant overestimates of X-ray cluster
masses. Our interpretation is supported by the small cluster gas mass
fraction implied by the X-ray estimates of total masses, with
$f_{\rm gas}(<r_{500})=0.060\pm0.014$, as compared to 
$f_{\rm gas}(<r_{500})=0.158\pm0.038$ using the lensing estimate of the
total mass. A merger along the line-of-sight can also explain the high
concentration parameter value found for the cluster, and is consistent
with the large velocity difference of  $\Delta v\simeq 2800~{\rm
  km\,s^{-1}}$ between the bright  central galaxies G1 and G2
\citep{oguri08}. The merger interpretation for this strong lens
selected cluster, in turn, may imply a large impact of cluster mergers
on strong lensing cross sections \cite[e.g.,][]{torri04,fedeli06,redlich12}.  

If the merger interpretation is correct, one question is whether the
large velocity difference is compatible with the theoretical
expectations. There are several works examining the distribution
of velocity differences between close halo pairs 
\citep[e.g.,][]{hayashi06,lee10,thompson12}, which indicate that a
relative velocity of $\Delta v\simeq 2800~{\rm km\,s^{-1}}$  for
haloes with $M_{\rm vir}\ga 10^{14}h^{-1}M_\odot$ is quite rare.
Therefore, it is important to make more quantitative comparisons 
taking account of the redshift and mass of SDSS J1029+2623 in order
to check whether this large velocity difference can be explained in
the framework of the standard $\Lambda$-dominated cold dark matter
cosmology. 

We confirm the earlier claims \citep{oguri08,kratzer11,ota12} that the
observed flux ratios between the quasar images cannot be reproduced by
strong lens models without any substructure in the vicinity of the
quasar images. From a careful inspection of the {\it HST} images, we
have identified a probable galaxy GX which is only $\sim 0\farcs4$
from quasar image C. A mass model including GX successfully reproduces
the observed flux ratios, thereby resolving the flux ratio
anomaly. While GX has little effect on the AB time delay measured by
\citet{fohlmeister12}, the BC time delay is a sensitive probe of its
effects. We have also checked the morphology of the lensed host galaxy
and found a very good agreement with the model prediction. The
best-fit model predict the total magnification of the quasar as
$\mu_{\rm tot}=30$ and that of the lensed quasar host galaxy as 
$\mu_{\rm tot}=35$.  

Finally we note that additional observations are essential to better
characterise this unique quasar-cluster lens system. For instance, the
measurement of the time delay between quasar images B and C will be
important for constraining the properties of galaxy GX. The merger
interpretation for the lensing cluster should be checked by extensive
spectroscopy of cluster member galaxies. Spectroscopy of the multiply
imaged galaxies identified in this paper is important both for more
robust mass models and for possible cosmological constraints 
\citep[e.g.,][]{jullo10}. Furthermore, the
highly elongated host galaxy makes  this lens system an ideal site to
conduct detailed explorations of a quasar host galaxy at $z\sim 2$
\citep[e.g.,][]{ross09}.   

\section*{Acknowledgments}
We thank an anonymous referee for useful suggestions. 
This work was supported in part by the FIRST program
``Subaru Measurements of Images and Redshifts (SuMIRe)'', World
Premier International Research Center Initiative (WPI Initiative),
MEXT, Japan, and Grant-in-Aid for Scientific Research from the JSPS 
(23740161). 
TS acknowledges support from NSF through grant AST-0444059-001, and
the Smithsonian Astrophysics Observatory through grant GO0-11147A.
CSK is supported by NSF grant AST-1009756.
XD acknowledges financial support from NASA/SAO award GO011147B and
NASA/STScI award HST-GO-12915.07-A. 
This paper is based on observations with the NASA/ESA Hubble Space
Telescope, obtained at the Space Telescope Science Institute,
which is operated by the Association of Universities for Research
in Astronomy, Inc., under NASA contract NAS 5-26555. 
This research has made use of data obtained from the Chandra Data
Archive and the Chandra Source Catalog, and software provided by the
Chandra X-ray Center (CXC) in the application packages CIAO, ChIPS,
and Sherpa. 
This paper is based on data collected at Subaru Telescope, which is
operated by the National Astronomical Observatory of Japan. 


\label{lastpage}


\begin{thebibliography}{}

\bibitem[Akamatsu et al.(2011)]{akamatsu11}
Akamatsu H., Hoshino A., Ishisaki Y., 
Ohashi T., Sato K., Takei Y., Ota N., 2011, PASJ, 63, 1019 

\bibitem[Allen et al.(2008)]{allen08} 
Allen S.~W., Rapetti D.~A., Schmidt R.~W., Ebeling H., 
Morris R.~G., Fabian A.~C., 2008, MNRAS, 383, 879 

\bibitem[Bar-Kana(1996)]{barkana96} 
Bar-Kana R., 1996, ApJ, 468, 17 

\bibitem[Bertin \& Arnouts(1996)]{bertin96}
Bertin E., Arnouts S., 1996, A\&AS, 117, 393 

\bibitem[Brammer, van Dokkum, \& Coppi(2008)]{brammer08}
Brammer G.~B., van Dokkum P.~G., Coppi P., 2008, ApJ, 686, 1503 

\bibitem[Broadhurst et al.(2005)]{broadhurst05}
Broadhurst T., Takada M., Umetsu K., Kong X., 
Arimoto N., Chiba M., Futamase T., 2005, ApJ, 619, L143 

\bibitem[Burns et al.(2010)]{burns10}
Burns J.~O., Skillman S.~W., O'Shea B.~W., 2010, ApJ, 721, 1105 

\bibitem[Coe et al.(2006)]{coe06} 
Coe D., Ben{\'{\i}}tez N., S{\'a}nchez S.~F., Jee M., Bouwens R., Ford H., 
2006, AJ, 132, 926 

\bibitem[Cohn et al.(2001)]{cohn01} 
Cohn J.~D., Kochanek C.~S., McLeod B.~A., Keeton C.~R., 2001, 
ApJ, 554, 1216 

\bibitem[Dai, Kochanek, \& Morgan(2007)]{dai07}
Dai X., Kochanek C.~S., Morgan N.~D., 2007, ApJ, 658, 917 

\bibitem[Dai et al.(2010a)]{dai10a}
Dai X., Kochanek C.~S., Chartas G., Koz{\l}owski S., Morgan C.~W., Garmire 
G., Agol E., 2010a, ApJ, 709, 278 

\bibitem[Dai et al.(2010b)]{dai10b} 
Dai X., Bregman J.~N., Kochanek C.~S., Rasia E., 2010b, ApJ, 719, 119 

\bibitem[Duffy et al.(2010)]{duffy10}
Duffy A.~R., et al., 2010, MNRAS, 405, 2161 

\bibitem[Ettori et al.(2009)]{ettori09}
Ettori S., Morandi A., Tozzi P., Balestra I., Borgani S., Rosati P.,
Lovisari L., Terenziani F., 2009, A\&A, 501, 61  

\bibitem[Fabjan et al.(2011)]{fabjan11} 
Fabjan D., Borgani S., Rasia E., Bonafede A., Dolag K., 
Murante G., Tornatore L., 2011, MNRAS, 416, 801 

\bibitem[Fedeli et al.(2006)]{fedeli06}
Fedeli C., Meneghetti M., Bartelmann M., Dolag K., Moscardini L.,
2006, A\&A, 447, 419  

\bibitem[Fedeli(2012)]{fedeli12}
Fedeli C., 2012, MNRAS, 424, 1244 

\bibitem[Fohlmeister et al.(2007)]{fohlmeister07}
Fohlmeister J., et al., 2007, ApJ, 662, 62 

\bibitem[Fohlmeister et al.(2012)]{fohlmeister12}
Fohlmeister J., Kochanek C.~S., Falco E.~E., 
Wambsganss J., Oguri M., Dai X., 2012, arXiv:1207.5776 

\bibitem[Hayashi \& White(2006)]{hayashi06}
Hayashi E., White S.~D.~M., 2006, MNRAS, 370, L38 

\bibitem[Inada et al.(2003)]{inada03} 
Inada N., et al., 2003, Natur, 426, 810 

\bibitem[Inada et al.(2006)]{inada06} 
Inada N., et al., 2006, ApJ, 653, L97 

\bibitem[Jee et al.(2011)]{jee11} 
Jee M.~J., et al., 2011, ApJ, 737, 59 

\bibitem[Jullo et al.(2007)]{jullo07} 
Jullo E., Kneib J.-P., Limousin M., El{\'{\i}}asd{\'o}ttir {\'A}., 
Marshall P.~J., Verdugo T., 2007, NJPh, 9, 447 

\bibitem[Jullo et al.(2010)]{jullo10} 
Jullo E., Natarajan P., Kneib J.-P., D'Aloisio A., 
Limousin M., Richard J., 
Schimd C., 2010, Sci, 329, 924 

\bibitem[Kaiser \& Squires(1993)]{kaiser93}
Kaiser N., Squires G., 1993, ApJ, 404, 441 

\bibitem[Kaiser et al.(1995)]{kaiser95}
Kaiser N., Squires G., Broadhurst T., 1995, ApJ, 449, 460 

\bibitem[Keeton \& Moustakas(2009)]{keeton09}
Keeton C.~R., Moustakas L.~A., 2009, ApJ, 699, 1720 

\bibitem[King \& Corless(2007)]{king07}
King L., Corless V., 2007, MNRAS, 374, L37 

\bibitem[Koekemoer et al.(2002)]{koekemoer02}
Koekemoer A.~M., Fruchter A.~S., Hook 
R.~N., Hack W., 2002, hstc.conf, 337 

\bibitem[Komatsu et al.(2011)]{komatsu11}
Komatsu E., et al., 2011, ApJS, 192, 18 

\bibitem[Kratzer et al.(2011)]{kratzer11}
Kratzer R.~M., Richards G.~T., Goldberg D.~M., 
Oguri M., Kochanek C.~S., Hodge J.~A., Becker R.~H., 
Inada N., 2011, ApJ, 728, L18 

\bibitem[Lee \& Komatsu(2010)]{lee10}
Lee J., Komatsu E., 2010, ApJ, 718, 60 

\bibitem[Li et al.(2007)]{li07}
Li G.~L., Mao S., Jing Y.~P., Lin W.~P., Oguri M., 2007, 
MNRAS, 378, 469 

\bibitem[Mahdavi et al.(2008)]{mahdavi08}
Mahdavi A., Hoekstra H., Babul A., Henry J.~P., 2008, 
MNRAS, 384, 1567 

\bibitem[McCarthy et al.(2010)]{mccarthy10}
McCarthy I.~G., et al., 2010, MNRAS, 406, 822 

\bibitem[Medezinski et al.(2007)]{medezinski07}
Medezinski E., et al., 2007, ApJ, 663, 717 

\bibitem[Medezinski et al.(2010)]{medezinski10}
Medezinski E., Broadhurst T., Umetsu K., Oguri M., Rephaeli Y., 
Ben{\'{\i}}tez N., 2010, MNRAS, 405, 257 

\bibitem[Meneghetti et al.(2007)]{meneghetti07}
Meneghetti M., Argazzi R., Pace F., Moscardini L., Dolag K.,
Bartelmann M., Li G., Oguri M., 2007, A\&A, 461, 25  

\bibitem[Minor \& Kaplinghat(2008)]{minor08}
Minor Q.~E., Kaplinghat M., 2008, MNRAS, 391, 653 

\bibitem[Miyazaki et al.(2002)]{miyazaki02}
Miyazaki S., et al., 2002, PASJ, 54, 833 

\bibitem[Motta et al.(2012)]{motta12} 
Motta V., Mediavilla E., Falco E., Munoz J.~A., 2012, 
ApJ, 755, 82 

\bibitem[Navarro, Frenk, \& White(1997)]{navarro97}
Navarro J.~F., Frenk C.~S., White S.~D.~M., 1997, ApJ, 490, 493 

\bibitem[Nelson et al.(2012)]{nelson12} 
Nelson K., Rudd D.~H., Shaw L., Nagai D., 2012, ApJ, 751, 121 

\bibitem[Oguri(2007)]{oguri07}
Oguri M., 2007, ApJ, 660, 1 

\bibitem[Oguri(2010)]{oguri10b}
Oguri M., 2010, PASJ, 62, 1017 

\bibitem[Oguri \& Keeton(2004)]{oguri04}
Oguri M., Keeton C.~R., 2004, ApJ, 610, 663 

\bibitem[Oguri et al.(2010)]{oguri10a} 
Oguri M., Takada M., Okabe N., Smith G.~P., 2010, MNRAS, 405, 2215 

\bibitem[Oguri et al.(2005)]{oguri05} 
Oguri M., Takada M., Umetsu K., Broadhurst T., 2005, ApJ, 632, 841 

\bibitem[Oguri et al.(2012)]{oguri12} 
Oguri M., Bayliss M.~B., Dahle H., Sharon K., Gladders M.~D., Natarajan P., 
Hennawi J.~F., Koester B.~P., 2012, MNRAS, 420, 3213 

\bibitem[Oguri et al.(2008)]{oguri08} 
Oguri M., et al., 2008, ApJ, 676, L1 

\bibitem[Okabe \& Umetsu(2008)]{okabe08}
Okabe N., Umetsu K., 2008, PASJ, 60, 345 

\bibitem[Okabe et al.(2010)]{okabe10} 
Okabe N., Zhang Y.-Y., Finoguenov A., Takada M., Smith G.~P., Umetsu K., 
Futamase T., 2010, ApJ, 721, 875 

\bibitem[Okabe et al.(2011)]{okabe11} 
Okabe N., Bourdin H., Mazzotta P., Maurogordato S., 2011, 
ApJ, 741, 116 

\bibitem[Ota et al.(2012)]{ota12} 
Ota N., et al., 2012, ApJ, 758, 26

\bibitem[Peng et al.(2002)]{peng02} 
Peng C.~Y., Ho L.~C., Impey C.~D., Rix H.-W., 2002, 
AJ, 124, 266 

\bibitem[Peng et al.(2006)]{peng06} 
Peng C.~Y., Impey C.~D., Rix H.-W., Kochanek C.~S., Keeton C.~R., Falco 
E.~E., Leh{\'a}r J., McLeod B.~A., 2006, ApJ, 649, 616 

\bibitem[Pratt et al.(2010)]{pratt10}
Pratt G.~W., et al., 2010, A\&A, 511, A85 

\bibitem[Rasia et al.(2011)]{rasia11} 
Rasia E., Mazzotta P., Evrard A., Markevitch M., Dolag K., 
Meneghetti M., 2011, ApJ, 729, 45 

\bibitem[Redlich et al.(2012)]{redlich12}
Redlich M., Bartelmann M., Waizmann J.-C., 
Fedeli C., 2012, A\&A, in press (arXiv:1205.6906)

\bibitem[Ricker \& Sarazin(2001)]{ricker01}
Ricker P.~M., Sarazin C.~L., 2001, ApJ, 561, 621 

\bibitem[Ross et al.(2009)]{ross09} 
Ross N.~R., Assef R.~J., Kochanek C.~S., Falco E., 
Poindexter S.~D., 2009, ApJ, 702, 472 

\bibitem[Schrabback et al.(2010)]{schrabback10}
Schrabback T., et al., 2010, A\&A, 516, A63 

\bibitem[Soucail(2012)]{soucail12}
Soucail G., 2012, A\&A, 540, A61 

\bibitem[Takizawa, Nagino, \& Matsushita(2010)]{takizawa10}
Takizawa M., Nagino R., Matsushita K., 2010, PASJ, 62, 951 

\bibitem[Thompson \& Nagamine(2012)]{thompson12}
Thompson R., Nagamine K., 2012, MNRAS, 419, 3560 

\bibitem[Torri et al.(2004)]{torri04} 
Torri E., Meneghetti M., Bartelmann M., Moscardini L., Rasia E., 
Tormen G., 2004, MNRAS, 349, 476 

\bibitem[Umetsu \& Broadhurst(2008)]{umetsu08}
Umetsu K., Broadhurst T., 2008, ApJ, 684, 177 

\bibitem[Umetsu et al.(2009)]{umetsu09} 
Umetsu K., et al., 2009, ApJ, 694, 1643 

\bibitem[Vikhlinin et al.(2006)]{vikhlinin06}
Vikhlinin A., Kravtsov A., Forman W., Jones C., 
Markevitch M., Murray S.~S., Van Speybroeck L., 2006, ApJ, 640, 691 

\bibitem[Zhang et al.(2010)]{zhang10} 
Zhang Y.-Y., et al., 2010, ApJ, 711, 1033 

\end{thebibliography}
\end{document}